\documentclass[11pt,prd,showpacs,showkeys]{revtex4}

\usepackage{mathrsfs,amsbsy,amssymb,latexsym,amsfonts,amsmath}
\usepackage[dvipdfm]{graphicx,color}
\usepackage{wrapfig}
\usepackage{amsbsy}
\usepackage{alltt}

\makeatletter
	
	\@addtoreset{equation}{section}
\makeatother

\begin{document}

\begin{flushright}
OIQP-12-13
\end{flushright}

\title{{Our String Field Theory
Liberating Left and Right Movers \\ as 
Constituent ``Objects''}}

\author{Holger B. {\sc Nielsen}}
\affiliation{Niels Bohr Institute, University of Copenhagen,\footnote[0]{To appear in the Proceedings of the 15th Workshop 
``What comes beyond the Standard Models'', Bled,
July 2012 eds. Norma M. Borstnik,
Holger B. Nielsen and D. Lukman.}\footnote{email: hbech@nbi.dk, hbechnbi@gmail.com} \\
Copenhagen $\phi$, DK2100, Denmark.}

\author{Masao {\sc Ninomiya}}
\affiliation{Okayama Institute for Quantum Physics, Kyoyama 1-9-1, Okayama 700-0015, Japan.\footnote{email:msninomiya@gmail.com }}

\begin{abstract}

\vspace{0.2cm}

\begin{center}
{\bf abstract}
\end{center} 

\vspace{-0.2cm}

We review the idea of our earlier proposed
string field theory
\cite{early1,early2,ourappear}, which 
makes the second quantized string 
theory appear as
described by one or two types of 
stationary - so called - 
``objects'' for string theories 
respectively with and without open 
strings. It may be better to 
look on our
string field theory as a {\em solution}
of a second quantized string theory, in 
which we have decided to ignore, how 
strings are topologically hanging 
together. Rather we satisfy 
ourselves with realizing 
solely the information contained in the 
knowledge of, through which points in 
space time passes a string. In the formulation 
of the string field theory, in which we
have rewritten the systems of strings 
into a system of what we call ``objects'',
the scattering of strings take place 
without any of the ``fundamental'' 
``objects'' (technically ``even objects'')
changing. They are only {\em exchanged}
instead. A route to extract from our 
formalism the vertex of the 
Veneziano model theory is sketched, and 
thus in principle we have a line of 
arguments leading to, that our string 
field theory indeed gives the 
Veneziano model scattering amplitudes for 
the scatterings, that are in fact in our 
model {\em only exchanges}.     
\end{abstract}



%\begin{tabular}{ll}
%\end{tabular}

%\thispagestyle{empty}
%\setcounter{page}{0}
%\tableofcontents

\maketitle

\section{Introduction}
In our earlier~\cite{early2,early1}  
and in a coming work ~\cite{ourappear}
we have presented a string field theory 
idea, the characteristic feature of 
which is, that it is based on the 
observation, that the left and right 
moving wave patterns on the strings are 
conserved according to a certain 
conservation theorem to be explained. 
Really that means,
that there are so many conservation laws, 
that we may consider our picture a 
{\em solution} of string theory. 
Although to some extent, if our 
formulation 
works as we suggest and almost prove,
it should represent  in some way the 
same string field 
theory as the other string field theories
on the market \cite{lcgbosonic, WittenSFT,closedWitten,JSWittenopen,JSWittenclosed,modifiedcubicWitten,susyWittenopen,susyGS,
reviewSFT} 
%~\cite{3}-\cite{11}
, there 
is {\em 
one important difference} in as far as
in our picture some information 
considered physical in the other models
 is considered not physical
and thus {\em ignored} in ours! In fact we 
consider the question of how various 
pieces  of strings are connected to 
each other as 
being {\em not a physical question}. 
Rather one can in our string field 
theory only discern the way in which
various pieces of strings
hang together from the continuity
(as function of the string parametrizing
parameter $\sigma$)
of the position(or momentum) variables.
If thus a couple of strings happen to 
have a common point, our formulation 
truly lacks an information compared to
the usual
\cite{lcgbosonic, WittenSFT,closedWitten,
JSWittenopen,JSWittenclosed,modifiedcubicWitten,susyWittenopen,susyGS,reviewSFT} 
%~\cite{3}-\cite{11} 
type of 
string field theories. From our point 
of view we might 
think that the older string field 
theories
\cite{lcgbosonic, WittenSFT,closedWitten,JSWittenopen,JSWittenclosed,modifiedcubicWitten,susyWittenopen,susyGS,reviewSFT} 
%~\cite{3}-\cite{11} are more 
are complicated, because they
{\em keep} this information about 
the hanging together of the string pieces. 
\begin{figure}[!htb]
 \includegraphics[width=0.6\textwidth]{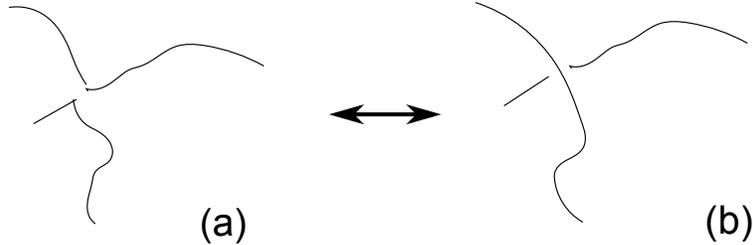}%
  \caption{In the very moment a scattering takes place for here a couple of open strings except that the couple of strings get split
           and the pieces united in a different way. But wherever there is some string bit it remains on both Figures 1(a) 
and 1(b). \label{Figure3}}
\end{figure}
Figure 1 illustrates the type of 
distinction, which we ignore in our model.

 In that sense the conventional theories 
seek to keep track of some hanging 
together information - which we ignore -
and there have to have a relative to 
our theory a complication by having to
store some way that - in our point 
of view - {\em extra} information 
in the formalism of the ``Fock space''.

But since this information is almost 
already contained in the continuity,
it is only when two strings pass through 
the 
{\em same} point that it gets really 
important,
and since this has  zero-probability for
happening, we may claim that at the end 
the important deviation is only on a 
null set of configurations! But there 
{\em is} a difference. And that fact 
of course ensures that our model is at 
least {\em new} in literature.

Also it means that our development of 
our string field theory is not built 
on the previous string field theories, 
but rather takes its outset alone from
string theory, almost as going back to its
initiation\cite{stringNielsen,stringNambu1,
stringNambu2,stringSusskind, Birth}. 
Really we basically start from the single
string description and only use string 
field theory in the sense, that we - 
in words - have in mind that there are 
several strings present at the same time.

To make sure that our string field theory,
as we shall describe it, in its terms
of so called ``objects''(which are 
mathematically related to the strings in a 
slightly complicated way to be described 
below or found in our earlier works on 
this string field theory\cite{early2,
early1,ourappear}) is indeed essentially
(ignoring some null sets) equivalent to 
usual string theory and thus also to 
the older string field theories 
\cite{stringNielsen,stringNambu1,
stringNambu2,stringSusskind, Birth}
we would like to show that it leads 
to the Veneziano model \cite{Veneziano}
and the usual string spectrum. The latter
we have shown in the coming up work
\cite{ourappear}, but it is not so
trivial again as one might at first think,
because we have the single string theory
as our outset; the point namely is to 
show that we out of our 
``object''-formalism effectively can 
deduce sufficiently much of the single 
string theory so as obtain this spectrum.

In the present article it is the major 
goal to come  close to deducing, that our 
``object''-formalism describing 
potentially an 
arbitrary number of strings and thus 
our ``string field theory'' leads to the Veneziano model 
amplitudes,
of course via getting rewritten into 
the derivation of Veneziano amplitudes 
from single-string  or rather a few 
strings theory.
%In section\ref{exchange} we illustrate 
%by a slightly oversimplified analogy 
%to a an infinitely loosely bound state 
%scattering during which the constituents
%do not interact, so that the scattering 
%truly gets very formal only.  
In the next section \ref{setup} we shall 
review our string field theory by 
sketching the connection between our 
``object''-Fock space and string theory
with several strings.     
In section\ref{exchange} we illustrate 
by a slightly oversimplified analogy 
to an infinitely loosely bound state 
scattering during which the constituents
do not interact, so that the scattering 
truly gets very formal only.
In the following section 
\ref{correspondence} we start developing 
a correspondence-formalism for a single 
open string and its corresponding 
close circular chain
 of ``objects''
call $F$. 
In section \ref{Fourier} we 
Fourier-transform - in the string 
$\sigma$-variable - the  
correspondence between strings and object
formulation for an open string. In this 
Fourier-transformed form  we
identify the usual creation and 
annihilation operators acting on 
the single string states in usual single
string formulation.
In section \ref{trickshort} we describe 
the calculational trick of going to a 
frame in which the plus-component of
one of three strings involved in the 
vertex, which we compute, goes to zero.
It is this trick that makes that string 
couple just to one point formally to the 
other strings considered otherwise to 
be just one of them becoming the other 
one. 
In the last section \ref{conclusion} we
conclude, that we have partly checked that
new ``object''-based ``string field 
theory'' represents string theory.

\section{Setting Up Our Model}
\label{setup}
We can naturally describe the model 
of ours in {\em two} opposite ways/orders,
since we can go either one way or the 
other.: 
\begin{itemize}
\item{{\bf objects $\rightarrow$ strings}}
That is to say we can first describe the 
``ontological picture'' of our model,
which consists of one if the model 
has ``even objects'' for which we
construct (a) Fock space(s). 
Next we then describe the mathematical
definitions by means of which we rewrite 
this/these Fock space(s) into a string 
theory - a string field theory really -
so that one can e.g. calculate scattering
amplitudes and thereby - we hope - obtain 
Veneziano amplitudes. 
\item{{\bf strings $\rightarrow$ objects}}
Alternatively we could start from the 
idea of having a string field theory 
in mind in the sense that we like, 
 Kaku and Kikkawa\cite{lcgbosonic} e.g. 
consider it 
that we have the possibility of having 
present any number of strings- even 
we can have a state being a superposition
of states with different numbers of 
strings present -. Then we rewrite
the set of strings present in terms first
of the right and left mover fields on the 
strings $X^{\mu}_R(\tau -\sigma)$ and 
$X^{\mu}_L(\tau -\sigma)$ - a set of 
26 functions for each string (two sets for 
a closed string) and next the derivatives
of these functions are represented at
the end by our ``objects''.
   \end{itemize}  

\subsection{The String to 
``Objects'' Way} 
Let us first describe the second way - 
from strings to objects - a bit more in 
detail: 

We consider a string field theory state
in principle, if we just think of a set 
of strings in various states. Let us - 
to be definite- call the number of strings
present $\Xi$ and enumerate a  
 string among these $\Xi$ ones by 
($\iota$
or $\kappa$ = 1, 2,...,$\Xi$.)
and then in the usual way we split the 
solution for the single string equations 
of motion 
\begin{equation}
(\partial_{\tau}^2 - \partial_{\sigma}^2)
X^{\iota\mu}(\tau,\sigma) =0 \hbox{    (
$\iota$=1,2,3,...,$\Xi$ ) }
\end{equation} 
into right and left movers - at least 
locally - 
\begin{equation}
X^{\iota\mu}(\tau, \sigma) = 
X_R^{\iota\mu}(\tau 
-\sigma) + X_L^{\iota\mu}(\tau +\sigma).
\end{equation}

The splitting into the left $X_L^{\iota\mu}
(\tau-\sigma)$ and the right mover part
$X_R^{\iota\mu}(\tau+\sigma)$ is a bit 
ambiguous in as far as a constant 
could be moved from left to right
or opposite. This ambiguity, however,
disappears, if we instead consider the 
derivatives with respect to of these 
left and 
right mover fields on the strings,
$\dot{X}_R^{\iota\mu}(\tau_{\iota} 
-\sigma_{\iota})$
and   $\dot{X}_R^{\iota\mu}(\tau -\sigma)$.
(Here the dot denotes derivative with 
respect to
say $\tau_{\iota}$).  

Now we shall have in mind that, while 
for say a closed string the right mover
derivative field $\dot{X}_R^{\iota\mu}(\tau
_{\iota} - \sigma_{\iota})$ commute with 
the left mover derivative field for string
$\iota$ say, then  e.g. the right mover 
fields 
derivative for various values of the 
argument $\tau_{\iota R}=
\tau_{\iota} -\sigma_{\iota}$
do {\em not} commute. In fact we 
have the commutation relation for these 
derivatives of the form
\begin{eqnarray}
\bigl[ \dot{X}^{\iota\mu}_{R}(\tau_{R}), 
\dot{X}^{\kappa\nu}_{R}(\tau_{R} ') \bigr]
= i2\pi \alpha' \delta^{\mu\nu}
\delta^{\iota\kappa} 
\frac{\partial}{\partial \tau_{R}} 
\delta(\tau_{R}'-\tau_{R}).
\label{commutationrule}
\end{eqnarray}

Now the main idea of our string field 
theory is to notice that thinking 
classically the {\em set} of vectorial 
values 
in the 25+1 dimensional Minkowski space 
of e.g. $\dot{X}_R^{\iota\mu}(\tau_{\iota}-
\sigma_{\iota})$ does {\em not} change as
time - let us hereby at least crudely
think of $\tau_{\iota}$ as a time - passes 
by, because of  $\dot{X}_R^{\iota\mu}
(\tau_{\iota}-\sigma_{\iota})$ only depends
on the combinations $\tau_{\iota} 
-\sigma_{\iota}$. Then namely a change in 
the ``time'' $\tau_{\iota}$ can be replaced 
by a corresponding shift in $\sigma_{\iota}$
and in that way ensure that for some 
$\sigma_{\iota}$ the same values in the 
25+1 dimensional Minkowski 
%space will be 
that are 
achieved for one $\tau_{\iota}$-value  
will also be achieved for another 
$\tau_{\iota}$-value. In that way the set 
of values taken on remains the same:
it is constant in ``time''. Because of 
this fact
we can think of the string $\iota$
as described by its right and left 
set of mover derivative vectorial 
values. Well, these say two sets 
of right and left movers - to take 
the closed string case - do not 
100 \% though describe the state of the 
string $\iota$, but it is very close to
be so. In fact there is of course 
an ambiguity with respect to adding a 
constant
to $X_R^{\iota\mu}(\tau_{\iota} 
- \sigma_{\iota})$ and another constant
to $X_R^{\iota\mu}(\tau_{\iota} 
- \sigma_{\iota})$, which in turn would
leave an ambiguity of adding a constant
to the position of the string. 

The truly important point is that 
classically thinking, {\em even when some 
strings scatter} in the way that some 
pieces of the incoming strings get 
distributed in new combinations forming 
the outgoing strings the system of 
vectorial values taking on by the 
derivatives $\dot{X}_{\iota'}^{\iota'\mu}
(\tau_{\iota'} - \sigma_{\iota'})$ for the 
various 
strings $\iota'$ does {\em not} have time to 
{\em change} in the in principle 
infinitely short
moments of ``time'' when some of the 

strings just touch each other. The 
conclusion from these remarks is the 
theorem(s) about the ``images'' of the 
derivatives $\dot{X}_{R}^{\iota'\mu}$
and  $\dot{X}_{L}^{\iota'\mu}$ in the 
Minkowski space of 25 +1 dimension(s)
saying that these ``images'' are preserved 
in time.
\cite{early1,early2, ourappear}.

The crux of the matter is that it is as if
there are some ``objects'' corresponding
to any vectorial value that is taken on by
these $\dot{X}_{R}^{\iota'\mu}$ or 
$\dot{X}_{L}^{\iota'\mu}$ and such ``objects''
do not appear or disappear neither as 
time goes on while the isolated strings 
just develop, nor while the strings even
scatter classically thinking. 

This then opens up the possibility of 
describing the {\em whole situation} - the 
whole state of the multistring system -
by means of these ``objects''.
That is to say: we let the state of the 
system of strings  -  to be a string field 
theory ``Fock''-like state - be described
by these ``objects'' meaning really the
vectorial values taken on by the 
derivative variables 
$\dot{X}_{R}^{\iota'\mu}$ and 
$\dot{X}_{L}^{\iota'\mu}$.
If we do so, you should see, that 
we obtain that the state of the ``objects''
will not change according to the 
introductory discussion - at least up to
the problems with the additive constants
in e.g. the string positions.

This opens the possibility of describing
the string field theory ``Fock''-like 
state by the in time not developing 
``objects''. That is to say that we have 
now found a way to describe the 
``Fock''-like states as {\em non-moving}. 
 
\subsection{Discretization and Even 
/ Odd Story}

In order to get a more precise formalism
we must imagine to {\em discretize } the 
variables $\tau_{R\iota} = \tau_{\iota} 
- \sigma_{\iota}$ and $\tau_{L\iota} 
= \tau_{\iota} + \sigma_{\iota}$  into a 
series of discrete
points. There shall of course be some 
parameter $a$ which shall go to zero, 
and then in that limit $a \rightarrow 0$ 
the
continuum variables should be effectively 
reestablished. We should also have 
in mind that these variables are only
well defined, when the ``gauge'' of single
string theory has been fixed. This gauge 
 means that even after the gauge fixing 
that were already made before we even got 
to the formulation with d'Alembertian 
equation of motion by replacing say the 
Nambu action by the quadratic form
$\int [ (\partial_{\tau}X^{\mu})^2 -
(\partial_{\sigma}X^{\mu})^2] d\sigma 
d\tau$,  
there still {\em remains some 
reparametrization} of the 
variables $(\tau, \sigma)$ as being 
allowed(as ``gauge'' symmetry of the 
theory). In 
fact this
left over or remaining reparametrization 
- after the first gauge choice - 
is especially simple, if expressed in 
terms of the ``mover''-variables  
 $(\tau_R, \tau_L) = 
(\tau -\sigma, \tau+\sigma)$, in which 
formulation we obtain a new set of allowed
variables $(\tau'(\tau,\sigma), 
\sigma'(\tau,\sigma))$ very simply, when we
talk about 
\begin{eqnarray}
\tau_R' & = & \tau' - \sigma'\\
\tau_L' & = & \tau' + \sigma', 
\end{eqnarray}  
namely as a transformation
\begin{eqnarray}
\tau'_R&=& \tau'_R(\tau_R)\\
\tau_L' &=& \tau'_L(\tau_L).
\end{eqnarray}

We shall in the present article make 
use of what is called ``light cone gauge''
and consists in using infinite momentum
frame, i.e. the metric tensor 
 \begin{eqnarray}
g_{\mu\nu}= 
\begin{array}{c}
+ - \ \ \overbrace{ \hspace{6em}}^{24}\\
\left(
 \begin{array}{cccccc}
 0 & 1 & 0 &\cdots & \cdots& 0\\
 1 & 0 & 0 & \cdots & \cdots & 0\\
 0 & 0 & -1 & \cdots & \cdots & 0\\
 \vdots & \vdots & & \ddots & & \vdots\\
 0 & 0& \cdots &\cdots& -1 & 0\\
 0 & 0 & \cdots & \cdots & 0 & -1
 \end{array}
\right)
\end{array}
\begin{array}{c}
 \\ + \\ - \\ 
   \left. \begin{array}{l}
    \\ \\ \\ \\    
   \end{array}
    \right\} 24
\end{array}
\label{metric}
\end{eqnarray}

We can - and this is the ``light-cone gauge'' 
- imagine that we as a gauge 
choice insist on a certain fixed value 
for $\dot{X}_{\iota R}^+(\tau_{\iota R})$
(and also for the left 
$\dot{X}_{\iota L}^+(\tau_{\iota L})$
in closed string case). 

Now also remember that as result of the 
gauge invariance of the Nambu action 
\begin{equation}
S \propto \int \sqrt{{(X'\cdot\dot X)}^2
-
{\dot X}^2 * {X'}^2} d\sigma d\tau 
\end{equation} 
there appear {\em constraints} concerning 
the derivatives $X'$ and $\dot{X}$ of the
26-vector fields on the string $X=X^{\mu}$.
These constraints are well-known to be
\begin{eqnarray}
{X'}^2 + {\dot{X}}^2& =&0,\\
X'\cdot \dot{X} &=&0,
\end{eqnarray} 
which by trivial algebra written into
the $\dot{X_R} = \frac{1}{2} 
(\dot{X} -X')$ and $\dot{X_L} = 
\frac{1}{2}(\dot{X} + X')$ notation 
becomes 
\begin{eqnarray}
\dot{X_R}^2& = & 0, \\
\dot{X_L}^2&=&0.
\end{eqnarray}
These equations concern in a simple way
just our variables - the 
``mover''-variables - differentiated, and 
thus they lead to simple the rule, that 
the 
``objects'' lie on the light cone (in 25+1 
space time). 
This restriction removes one dimension
as degrees of freedom for the objects.
Since further as we saw the +component
is fixed by gauge choice, we have out of 
the original 26 = 25 +1 variables for 
each point of $\dot{X_R}$ (or for 
$\dot{X_L}$) or equivalently out of the 
``objects'' $J$ right or left {\em 
only $26 -1 -1= 24$ left over as truly
independent variables.}

We define the ``objects'' by putting
\begin{eqnarray}
J^{\mu}_L(\iota, I)& =&
\int_{\tau_{L\iota}(I -\frac{1}{2})}^
{\tau_{L\iota}(I+\frac{1}{2})} 
\dot{X_L}^{\mu}(\tau_L) 
d\tau_L,\\
J^{\mu}_R(\iota, I)& =&
\int_{{\tau_{L\iota}}(I -\frac{1}{2})}
^{\tau_{L\iota}(I+\frac{1}{2})} \dot{X_L}^{\mu}
(\tau_L) d\tau_L,
\end{eqnarray} 
and letting it be understood that the 
``separation'' values such as 
$\tau_{L\iota}(I -\frac{1}{2})$ and
$\tau_{L\iota}(I +\frac{1}{2})$ and the 
analogous ones for $R$ are to be 
constructed so that the +components 
indeed become simple constants 
proportional
to the discretization parameter $a$.

The restrictions on the $\dot{X_R}$ and 
$\dot{X_L}$ of course leads to the 
corresponding ones for the 
``object''-variables $J_R$ and $J_L$
(for small $a$ of course):
\begin{eqnarray}
J_R^{\mu}(\iota, I)J_R^{\nu}(\iota,I)
g_{\mu\nu} &=& 0,\\
J_L^{\mu}(\iota, I)J_L^{\nu}(\iota,I)
g_{\mu\nu} &=& 0,\\
J_R^+(\iota, I)= \frac{a\alpha'}{2},\\
J_L^+(\iota, I)= \frac{a\alpha'}{2}.
\end{eqnarray}
The only components, which are to be 
considered independent dynamical 
variables, are the 24 ``transverse'' 
components for which we denote the 
values of $\mu$ by $i$,where then
 $i=1, 2, ..., 23,24$.

These remaining variables  
 $J_R^i(\iota, I)$ and 
$J_L^i(\iota, I)$ for the ``objects''
may at first look like, that we 
could take them to be independent degrees
of freedom, but the non-zero commutation 
rules (\ref{commutationrule}) does not
allow to have the  different 
$J_R$-variables commute among each other.
It were  a major progress in the 
development of our string field theory 
project to find an idea of how to make 
the commutation rule 
(\ref{commutationrule})
get consistent with a discretization 
formulation. The trick is to
{\em let only every second - say the 
ones with even $I$ - of the $J_R^i(\iota,
I)$ and the $J_L^i(\iota,
I)$ be physical independent degrees 
of freedom.} Then the ``odd'' $I$ numbered
$J_R^i(\iota,I)$ and  $J_L^i(\iota,I)$
shall instead be {\em constructions 
made from the conjugate variables of the 
even ones}. In fact we shall take 
\begin{align}
	J^i(I)=-\pi \alpha^{\prime}\left
\{\Pi^i(I+1)-\Pi^i(I-1)\right\},
\end{align} 
where $\Pi^i(I)$ is the conjugate variables
to $J^i(I)$ - we shall of course 
imagine to put on either $R$ or $L$ 
in the closed string case. That is to say 
we have the commutation rule
\begin{equation}
[J^i(I), \Pi^k(K)] = i\delta^{ik}\delta_{IK}
\end{equation}
(where $K$ is an enumeration index of the
same type as $I$ and the index $R$ or $L$
in the only open string case as well as 
the string enumerating $\iota$ have been 
left out for simplicity)
which we a priori now only assume for 
even $I$ and $K$ in the philosophy that 
it is {\em only the even numbered 
``objects'' that truly exist} and thus 
shall be considered separate degrees 
of freedom. Really we only use the 
notation of $\Pi^i(I)$ for {\em even $I$}.

Notice that we have now come to describe 
a system of {\em arbitrarily} many 
strings by means of two (it turns out
that for theories with also open strings
 only a union of the right and the 
left objects shall be used instead of 
the two classes of objects for the 
only closed case.) sets of 
``even objects'' - meaning ``objects''
with even number enumerations - each 
having 24 variables $J_R^i$ or $J_L^i$
and their canonically conjugate variables
$\Pi^i_R$ and $\Pi_L^i$.

If we ignore quantum fluctuations of the 
``objects'', and if they come from 
continuous strings the objects coming 
e.g. from the right-mover $\dot{X_R}^{\mu}$
will lie on a continuous closed curve,
we can call it a ``cyclically ordered 
chain''. In fact the objects along such a 
chain is indeed ordered, so that each 
member - each object - has a successor.
This is true whether we consider only 
the even objects or include the odd ones 
too. Because of the continuity of the 
chain, you could essentially - i.e. 
with very little/few mistakes - deduce the 
ordering from the values of the 
$J_R^{\mu}$'s, so that delivering the 
{\em information about the ordering of the 
chains 
is almost superfluous}. 

A philosophical attitude like the 
following becomes possible because of 
this
containment of the information about the 
ordering already present in the values
of the $J$'s:

{\bf We have the freedom to declare
that the ordering is not a fundamentally
existing ``degree of freedom'' and thus 
of saying: There is no ordering 
information in the fundamental string 
field theory ``Fock space'', it has at 
the end to be extracted from the 
``object'' $J$ degrees of freedom alone,
by using the continuity.}

If we satisfy ourselves with this
observation saying, that we do not need the 
ordering information of the objects along
the chain corresponding to a string,
then we can imagine simplifying the 
string field theory to only have 
information about the - actually only
the even - separate ``objects''. 
This is a significant simplification 
and contributes significantly to make 
our string field theory very simple 
compared to competing string field 
theories. 

With this simplification - of 
throwing away the ordering in 
chain information - all that has to 
be described in the string field theory 
of ours is then: How many ``objects'' 
have a given combination of its
24 degrees of freedom $(J_R^i, \Pi_R^i)$
(and in the closed strings only case 
also $(J_L^i,\Pi_L^i)$)?  That can then 
be described by constructing a Fock-space
of the usual particle type, in which we 
construct a creation and annihilation 
operator for each value combination 
of say $J_R^i$. That is to say that 
we could e.g. decide to write the 
creation $a^+(J_R)$ and annihilation 
operators $a(J_R)$
as depending on the $J_R^i$-variables
taken only for the even objects. An 
alternative would be to write the 
creation $b^+(\Pi_R)$ and annihilation 
operators $b(\Pi_R)$ 
as functions of the $\Pi_R^i$ variables,
but as is well-known from ordinary quantum 
field theories you have to make a choice 
of one or of the other. This is 
 analogous to, that 
one in quantum field theory must choose 
either to write the creation and 
annihilation operators  as function of 
momentum 
$a^+(\vec{p})$ and $a(\vec{p})$
( and spin) or as function of the 
position variables ${\bf \phi}^+(x)$
(= the second quantized field)  and 
 ${\bf \phi}(x)$. 

Analogously to the expansions of the 
fields  ${\bf \phi}(x)$ in terms of
the momentum bases $a(\vec{p}) $ in 
quantum field theory we must have of 
course an analogous relation now 
of the form
\begin{equation}
b(\Pi_R) \propto \int a(J_R) 
\exp(i \sum_{i=1}^{24} \Pi_R^i J_R^i) d^{24}J.
\end{equation}       
Both the $a(J_R)$ and the $b(\Pi)$ 
should act on the same Fock space,
which now is in our model the Fock space 
for the second quantized string.
We should of course in the only closed 
string theory case have a Fock space 
which is a Cartesian product ${\cal H}_R 
\otimes {\cal H}_L$ of the one 
${\cal H}_R$ 
based on $J_R$ and $\Pi_R$ with an 
analogous one ${\cal H}_L$ based on the 
left mover variables $J_L$ and $\Pi_L$ 
instead.

In the also-open-string case, however, 
we get the mover-sets of variables mixed
up and could leave out the index $R$ 
or $L$. This comes about because at
the end of an open string the right mover
wave gets reflected as a left mover one
or oppositely.

One should notice that the formulation 
of a state of several strings in the 
formalism of these ``objects'' - or 
equivalently in terms of the string 
field theory Fock space of ours describing
the number of ``objects'' with a given 
value-set for $J_R$ say - there is no
time development, because the ``objects''
are static according to the above 
mentioned 
theorem of the non-variation of the 
images of the right and left mover 
$\dot{X_R}$ and $\dot{X_L}$ functions
into the 25+1 Minkowski spaces(where they
even only fall on the light cones).
In other words our Fock space describes
the world of strings at all times. Only 
the translation to the string language 
is needed, no time development calculation.
        
\subsection{The ``Object'' to String 
Way}

Let us now shortly contemplate the 
opposite way of describing our string 
field theory, namely to start from 
thinking on the ``objects'' described 
by their Fock space, and then look for
how one should get to translate that 
theory into a theory of a number of  
strings.

So imagine that we have a ``fundamental''
model (which should of course, if string 
theory 
were the theory of everything(``T.O.E.''),
 according to our picture be the 
fundamental theory
of nature)  described by a Hilbert
space or Fock space on which we have 
defined the operators $a_R(J_R)$ and 
$a_L(J_L)$ (and in addition one can 
define their Fourier transformed 
$b_R(\Pi_R)$ and $b_L(\Pi_L)$) operators 
annihilating ``objects'' with their 
24-component $J_R$ quantum numbers
being just 
$J_R = (J_R^1, J_R^2, ...,J_R^{24})$.
In the only-closed-string model we have 
also
corresponding $L$-operators, which we 
leave our for simplicity, or the reader 
may think 
of the open string case and ignore the 
indexes $R$ ( and $L$).  

Notice that our ``fundamental'' model
is really extremely {\em simple}: It is 
like 
a quantum field theory in which the 
particle has just 24 spatial dimensions
and nobody cares for any development in 
time. 

Then the story this way -i.e.{\em from
``objects'' to strings} - is a story 
about making mathematical definitions 
by means of the ``objects'' and then 
reach to {\em rewrite the state of the 
``objects'' into a state of some 
number of strings}. The string formulation
looks more complicated than our ``object''
formulation!

Let us here just enumerate the main steps,
and leave it for the reader to grasp
what shall go on by looking on the 
other way description above or by 
reading our other articles e.g. 
\cite{ourappear}:
\begin{itemize}
\item Interprete the Fock-space state 
of our ``fundamental'' model as
a quantum theory with an arbitrary 
number of ``objects'' (this is just what
one always does (analogously) in quantum 
field
theories)

\item Classify the ``objects'' into 
``cyclically ordered chains'' of 
``objects'',
so that these lie in approximately 
continuous closed chains. Although 
one could imagine working on making this
 step well-defined, it is a priori somewhat
ambiguous - and of course depends on 
``objects'' in states with quantum 
fluctuation. We must work more on this 
step, but the ambiguity may be important
for at all getting scattering in our 
string theory at the end.

\item Once we have these cyclically ordered
chains of (fundamental ~ even) ``objects''
we {\em invent } purely mathematically
``odd objects'', which by definition in the
now doubled cyclically ordered chains 
sit in between the original even or 
fundamental ``objects''. The odd 
``objects'' are of 
course having their $J_R$ constructed 
as differences of the conjugate variables
to the in the cyclically ordered chains 
neighboring even objects.
\item (Here is again a little trouble
to be worked on), but let us next 
pair a right and a left cyclically 
ordered chain and then construct at 
least the derivatives $\dot{X}$ and 
the $\sigma$-derivative $X'$ for all 
the combinations of a right and a left
``objects'' in the two chains which 
are being combined.
\item The strings determined at least 
essentially from these derivatives are 
now the strings corresponding to the 
Foch state we started from.  
\end{itemize}

It is to be more worked out in our model,
if the lack of getting the integration 
constant in the position of the strings
can be identified with the ambiguity of
choosing the origin of the 25+1 dimensional
Minkowski space, in which the strings are 
present.

With respect to at least the ``transverse''
momentum components, i.e. the ones with
$\mu=i=1,2,...,24$, it is easy to see 
that the even(or fundamental) ``objects''
function as constituents in the sense that
the ``transverse'' momentum of the string
is given as the sum over these even 
``object'' $J$-components. The odd 
``objects'' (the constructed ones) namely
cancel out in the sum giving the momentum,
because a certain $\Pi(I)$ contributes 
in two odd $J$'s with opposite signs,
and multiplied by the same $-\pi \alpha'$.
In this sense the even objects function 
much like constituents while then the 
string is the composed object the 
bound state. But it is a bit more 
complicated in detail.

Nevertheless we shall illustrate as 
an attempt to be pedagogical in the 
following describe an analogy to bound 
states and their constituents.

\section{Scattering by Just 
Exchange Is Thinkable}
% here I intend to essentially steal 
% the section 2 from ``A novel...''

%\section{Scattering without anything 
%happening \label{section 2}}  
\label{exchange}
In the ontological picture of our 
string field theory
 the Hilbert space of possible state of 
the world of a string theory 
 is in fact the Fock space for what we 
call ``objects''. These ``objects'' are each 
essentially a particle or a system with 
$24$ degrees of freedom, meaning 
$24 \ J^{i}(I)$-variables and 
$24 \ \Pi^{i}(I)$- variables canonically conjugate to the $J^{i}$'s.

Even when these ``objects'' in a 
slightly complicated way 
are rewritten to
describe 
scattering of strings, they themselves 
do \underline{not} develop even during 
the scattering. In our ``object'' - 
formulation everything is totally 
static, or rather there is no time, it 
is timeless. 
 
This scattering without anything changing 
sounds a priori very strange. Therefore 
we would like here to give at least an 
idea of how that strange phenomenon can 
come about:

Suppose that we had a couple of series 
of constituent particles making up some 
composite particles (essentially bound 
states, but they might not even be 
bound; rather just formally considered 
composed). Now if one suddenly decide to 
divide the ``constituent'' particles into 
groups forming composites in a different 
way from at first.

\begin{figure}[!htb]

 \includegraphics[width=0.35\textwidth]
{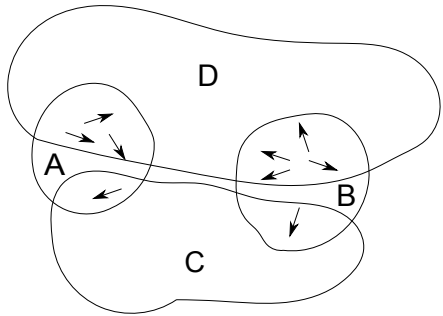}%
\caption{We illustrate how one may look at a set of (independent) ``constituents'' 
forming first two clumps $A$ and $B$, 
while later we divide them into two 
clumps $C$ and $D$ in a different way. 
\label{Figure2-a}}
\end{figure}

\begin{figure}[!htb]

 \includegraphics[width=0.35\textwidth]
{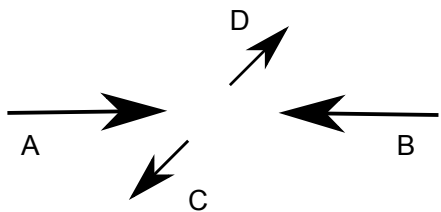}%
  \caption{Counting the momenta of the 
clump $A, B, C, D$ of the foregoing 
figure \ref{Figure2-a} as the sum of 
the momenta of the ``constituents'' 
           we get the picture of 
a \underline{scattering} $A+B \to C+D$.
\label{Figure2-b}}
\end{figure}

Then the momenta of the composite clumps 
after the considering the new ones will 
typically be quite different from those 
of the initial composite clumps.

This is a quite trivial remark: If we 
reclassify some constituents into a new 
set of classes of constituents (i.e. new 
composites) of course it will look like 
scattering of the composites.

In this way we hope that the reader can 
see that there is also a chance that 
making up a model on our ``objects'': 
these ``objects'' can function much like 
the just mentioned ``constituents'' and 
thus we could also in our model have some 
pretty formal scatterings. In the way we 
here think of the scatterings these 
scatterings are something we only think 
upon. The constituents and analogously 
our ``objects'' do not change their 
momenta, say, at all.  We have in this 
sense presented the idea of having 
completely formal scattering without 
anything going on at all for our 
``objects''. It is our hope in the long 
run to argue that in spite of this 
scattering being very formal we shall at 
the end get for it scattering amplitudes 
becoming the Veneziano model amplitudes.

%%%%%%%%%%%%%%%%%%%%%%%%%%%%%%%%%%%%%%%%%%%%%%%%%%%%%%%%%%%%%%%%%%%%%%%%%%%%%%%
\section{
Correspondence for Single (open) String}
\label{correspondence}
For a cyclically ordered chain of objects -- considering only the even ones (as existing) and only the ``transverse $J^i$ and $\Pi^i$" we have a wave function 
\begin{align}
	\psi (J^i(0),J^i(2),\cdots ,J^i(N-2))
\end{align}
(i.e. it is a wave function defined on a $24\cdot \frac{N}{2}$ dimensional space.)

Since such a chain represents an open 
string, this wave function should be essentially in corresponds with an open 
string state meaning a single string Fock 
space constructed from the creation and annihilation operators $a_n^i$ and $a_n^{i\dagger}=a_{-n}^i$. In the GGRT\cite{GGRT} we only have the transverse components $a_n^i, a_n^{i\dagger}$ with $i=1,2,\cdots ,24$. ``Essentially" here means that the $J^i(I)$, and $\Pi^i(I)$ I even gives us the $\dot{X}_R^i$ and $\dot{X}_L^i\simeq \dot{X}_R^i$ but we need an extra discussion to identify 
the ``average" position of the string associated with the cyclically ordered chain to be the average of $\Pi^i(I)$'s (which we postpone till later). Also the permutation of the ``objects" needs discussion.

Thus in principle there exists an operator $F$ mapping a wave function state $\psi$ for the cyclically ordered chain into the corresponding state $|\text{str }\psi \rangle$ of the open string in the conventional open string Fock space notation: 
\begin{align}
	|\text{str }\psi \rangle =F\psi .
\end{align}

One way to construct this correspondence $F$ could be to discretize the string analogously to the discretization, we used for $\dot{X}_R^{\mu}(\tau_R)=\dot{X}_L^{\mu}(\tau_R)$. This means that we consider the string at a specific ``time" $\tau$ (say $\tau =0$) and the discretize $\sigma$ in intervals so that $\int_{\text{an interval}}X_R^{\dagger}d\sigma =\frac{\alpha^{\prime}a}{2}$.

If we take the discretized steps in $\sigma$ to match with the discretization used for our objects it should mean that for each little discretized interval in $\sigma$, say $\Delta \sigma$ we obtain to say $\int_{\Delta \sigma}\dot{X}^{\mu}d\sigma$ comes from just \underline{two} (usually different) ``objects" -- one functioning as right-mover, the other one as left-mover --. The in $\sigma$ discretized string variables are then of the form 
\begin{align}
	\int_{\sigma -\Delta \sigma /2}^{\sigma +\Delta \sigma /2}
	X^{\mu \prime}(\sigma ,\tau)d\sigma 
	& =X^{\mu}\left(\sigma +\frac{\Delta \sigma}{2},\tau \right)
	-X^{\mu}\left(\sigma -\frac{\Delta \sigma}{2},\tau \right) \nonumber \\
	& =\int_{\sigma -\Delta \sigma /2}^{\sigma +\Delta \sigma /2}\left\{
	X_R^{\mu \prime}(\tau -\sigma)+X_L^{\mu \prime}(\tau +\sigma)\right\}
	d\sigma \nonumber \\
	& =-J^{\mu}(I)+J^{\mu}(K)
	\label{irsvc7a}
\end{align}
and
\begin{align}
	\int_{\sigma -\Delta \sigma /2}^{\sigma +\Delta \sigma /2}
	\dot{X}^{\mu}(\sigma ,\tau)d\sigma 
	& =\int_{\sigma -\Delta \sigma /2}^{\sigma +\Delta \sigma /2}
	\pi \alpha^{\prime}\Pi^{\mu}(\sigma ,\tau)d\sigma \nonumber \\
	& =J^{\mu}(I)+J^{\mu}(K)
	\label{irsvc7b}
\end{align}
where now if we assume say $I\simeq 0$ corresponds to $\sigma =0$ we have 
\begin{align}
	& I\simeq \frac{\sigma}{2\pi /N}=\frac{N\sigma}{2\pi}, \\
	& K\simeq -\frac{\sigma}{2\pi /N}=-\frac{N\sigma}{2\pi}.
\end{align}

Really we should not write our identification relations (\ref{irsvc7a},\ref{irsvc7b}) but rather we should include into them the operator $F$ going from the chain-Hilbert space to the string-one so that the two sides of the equations (\ref{irsvc7a},\ref{irsvc7b}) act on the same Hilbert space: 
\begin{align}
	F^{-1}\int_{\sigma -\Delta \sigma /2}^{\sigma +\Delta \sigma /2}
	X^{\mu \prime}(\sigma ,\tau)d\sigma F
	& =F^{-1}\left\{
	X^{\mu}\left(\sigma +\frac{\Delta \sigma}{2},\tau \right)
	-X^{\mu}\left(\sigma -\frac{\Delta \sigma}{2},\tau \right)\right\}F 
	\nonumber \\
	& =F^{-1}\int_{\sigma -\Delta \sigma /2}^{\sigma +\Delta \sigma /2}
	\left\{X^{\mu \prime}(\tau -\sigma)+X^{\mu \prime}(\tau +\sigma)
	\right\}d\sigma F \nonumber \\
	& =-J^{\mu}(I)+J^{\mu}(K) \nonumber \\
	& =-J^{\mu}\left(\frac{N\sigma}{2\pi}\right)
	+J^{\mu}\left(-\frac{N\sigma}{2\pi}\right).
\end{align}

%%%%%%%%%%%%%%%%%%%%%%%%%%%%%%%%%%%%%%%%%%%%%%%%%%%%%%%%%%%%%%%%%%%%%%%%%%%%%%%
\section{Fourier Transforming}
\label{Fourier}
We shall now seek the precise correspondence between the well-known $\alpha_n^{\mu}$ annihilation and for negative creation operators given in the notation of the formula (2.1.56) in the book by Green, Schwarz and Witten \cite{GSW} 
\begin{align}
	X^{\mu}(\sigma ,\tau)=x^{\mu}+l^2p^{\mu}\tau 
	+il\sum_{n\neq 0}\frac{e^{-in\tau}}{n}\alpha_n^{\mu}\cos (n\sigma)
	\label{GSWexpsvc4}
\end{align}
and the $c_L^{\mu}$ and $d_L^{\mu}$ in our previous article 
%\verb|
\cite{early2}
%|
%\textcolor{red}{$\Leftarrow$(undefined!!)} given by (63) and (64) 
%
\begin{align}
	c_L^i=\frac{1}{N}
	\sum_{\genfrac{}{}{0pt}{}{I=0,2,4,\cdots}{I\text{ even}}}^{N-2}
	J^i(I)e^{-\frac{i2\pi L\cdot I}{N}}
\end{align}
and 
\begin{align}
	d_L^i=\frac{1}{N}
	\sum_{\genfrac{}{}{0pt}{}{I=0,2,4,\cdots}{I\text{ even}}}^{N-2}
	\Pi^i(I)e^{-\frac{i2\pi L\cdot I}{N}}.
\end{align}
We can instead of these last two equations defining $c_L^i$ and $d_L^i$, namely 
(53-54), use the opposite Fourier transformations 
\begin{align}
	J^i(I) &=Re \left
\{\sum_{L=0}^{N-1}c_L^i
	\exp \left(\frac{iL\cdot I2\pi}{N}\right)\right\} \nonumber \\
	& =2Re \left\{\sum_{L=0}^{\frac{N}{2}-1}c_L^i
	\exp \left(\frac{iL\cdot I2\pi}{N}\right)\right\}
\end{align}
and (55) in 
%\verb|
\cite{early2}
%|
%\textcolor{red}{$\Leftarrow$(undefined!!)} 
%
\begin{align}
	\Pi^i(I) &=Re \left\{\sum_{L=0}^{N-1}d_L^i
	\exp \left(\frac{iL\cdot I2\pi}{N}\right)\right\} \nonumber \\
	& =Re \left\{\sum_{L=0}^{\frac{N}{2}-1}d_L^i
	\exp \left(\frac{iL\cdot I2\pi}{N}\right)\right\}
	\label{pidsvc13}
\end{align}
for even $I=0,2,\cdots ,N-4,N-2$ and $i=1,2,3,\cdots ,24$. Here the $c_L^i$'s and $d_L^i$'s obey 
\begin{align}
	c_L^i=c_{L+\frac{N}{2}}^i=\left(c_{-L}^i\right)^{\ast}
\end{align}
and
\begin{align}
	d_L^i=d_{L+\frac{N}{2}}^i=\left(d_{-L}^i\right)^{\ast}.
\end{align}
In (\ref{GSWexpsvc4}) the quantity 
\begin{align}
	l=\sqrt{2\alpha^{\prime}}=\frac{1}{\sqrt{\pi T}}
\end{align}
were just introduced to provide a 
length scale. Here $T$ is the string 
tension and $\alpha^{\prime}$ the Regge 
slope.

We may differentiate (\ref{GSWexpsvc4}) 
with respect to $\tau$ and insert the $l$-value 
$\sqrt{2\alpha^{\prime}}$ to obtain 
\begin{align}
	\dot{X}^{\mu}(\sigma ,\tau)=2\alpha^{\prime}p^{\mu}
	+\sqrt{2\alpha^{\prime}}
	\sum_{\genfrac{}{}{0pt}{}{n\neq 0}{-\infty}}^{\infty}e^{-in\tau}
	\alpha_n^{\mu}\cos (n\sigma).
	\label{dersvc14}
\end{align}
Similarly we could differentiate with respect to $\sigma$ to obtain 
\begin{align}
	X^{\mu \prime}(\sigma ,\tau)
	=-i\sqrt{2\alpha^{\prime}}
	\sum_{\genfrac{}{}{0pt}{}{n\neq 0}{-\infty}}^{\infty}e^{-in\tau}
	\alpha_n^{\mu}\sin (n\sigma).
	\label{dersvc15}
\end{align}
Remembering that locally 
\begin{align}
	X^{\mu}(\sigma ,\tau)=X_R^{\mu}(\tau -\sigma)+X_L^{\mu}(\tau +\sigma)
\end{align}
so that 
\begin{align}
	\dot{X}^{\mu}(\sigma ,\tau)
	=\dot{X}_R^{\mu}(\tau -\sigma)+\dot{X}_L^{\mu}(\tau +\sigma)
\end{align}
and 
\begin{align}
	X^{\mu \prime}(\sigma ,\tau)
	=-\dot{X}_R^{\mu}(\tau -\sigma)+\dot{X}_L^{\mu}(\tau +\sigma)
\end{align}
we obtain easily 
\begin{align}
	\dot{X}_L^{\mu}(\tau +\sigma)
	=\frac{1}{2}\left\{\dot{X}^{\mu}(\sigma ,\tau)
	+X^{\mu \prime}(\sigma ,\tau)\right\}
	\label{lsvc15a}
\end{align}
and 
\begin{align}
	\dot{X}_R^{\mu}(\tau -\sigma)
	=\frac{1}{2}\left\{\dot{X}^{\mu}(\sigma ,\tau)
	-X^{\mu \prime}(\sigma ,\tau)\right\}.
	\label{lsvc15b}
\end{align}
Inserting (\ref{dersvc14},\ref{dersvc15}) into these last two equations (\ref{lsvc15a},\ref{lsvc15b}) we obtain: 
\begin{align}
	\dot{X}_L^{\mu}(\tau +\sigma)
	& =\frac{1}{2}\left\{\dot{X}^{\mu}(\sigma ,\tau)
	+X^{\mu \prime}(\sigma ,\tau)\right\} \nonumber \\
	& =\sqrt{2\alpha^{\prime}}
	\sum_{\genfrac{}{}{0pt}{}{n\neq 0}{n=-\infty}}^{+\infty}
	\alpha_n^{\mu}\left\{\cos (n\sigma)-i\sin (n\sigma)\right\}e^{-in\tau}
	+\alpha^{\prime}p^{\mu} \nonumber \\
	& =\alpha^{\prime}p^{\mu}+\sqrt{2\alpha^{\prime}}
	\sum_{\genfrac{}{}{0pt}{}{n\neq 0}{n=-\infty}}^{+\infty}
	\alpha_n^{\mu}e^{-in\tau -in\sigma} \nonumber \\
	& =\alpha^{\prime}p^{\mu}+\sqrt{2\alpha^{\prime}}
	\sum_{\genfrac{}{}{0pt}{}{n\neq 0}{n=-\infty}}^{+\infty}
	\alpha_n^{\mu}e^{-in(\tau +\sigma)}
\end{align}
and 
\begin{align}
	\dot{X}_R^{\mu}(\tau -\sigma)
	& =\frac{1}{2}\left\{\dot{X}^{\mu}(\sigma ,\tau)
	-X^{\mu \prime}(\sigma ,\tau)\right\} \nonumber \\
	& =\alpha^{\prime}p^{\mu}+\sqrt{2\alpha^{\prime}}
	\sum_{\genfrac{}{}{0pt}{}{n\neq 0}{n=-\infty}}^{+\infty}
	\alpha_n^{\mu}e^{-in(\tau -\sigma)}.
\end{align}
If we put $\tau =0$ and discretize using $I=\frac{\sigma N}{2\pi}$ we may identify say 
\begin{align}
	e^{-in(\tau -\sigma)}=e^{+i\frac{2\pi IL}{N}}
	\label{idsvc17}
\end{align}
for the $\dot{X}_R$, where we took 
\begin{align}
	n=L
\end{align}
and for the $\dot{X}_L$ 
\begin{align}
	e^{-in(\tau +\sigma)=e^{i\frac{2\pi IL}{N}}}
\end{align}
also with $n=L$. Then we have of course 
\begin{align*}
	& \int \cdots d\sigma \Leftrightarrow 
	\sum_{I\text{ even}}\cdots \frac{2\pi}{N}\cdot 2. \\
	& \text{(the last factor $2$ from $I$ even only)}
\end{align*}
In this way we see comparing to 

\begin{align}
	J^i(I) & =\int_{\tau_R(I-\frac{1}{2})}^{\tau_R(I+\frac{1}{2})}
	\dot{X}_R^i(\tau_R)d\sigma \nonumber \\
	& =\frac{2\pi}{N}\cdot \dot{X}_R^i(\tau_R(I)) \nonumber \\
	& =2Re \left\{\sum_{L=0}^{\frac{N}{2}-1}c_L^i
	\exp \left(\frac{iL\cdot I2\pi}{N}\right)\right\}
\end{align}
that
\begin{align}
	\alpha^{\prime}p^i
+\sqrt{2\alpha^{\prime}}
	\sum_{\genfrac{}{}{0pt}{}{n\neq 0}{n=-\infty}}^{+\infty}
	\alpha_n^{\mu}e^{-in(\tau -\sigma)}=
%\end{align}
%
%\textcolor{red}{[original p.19 is 
%missing...]} 
%%% It is here I shall attempt to
%%%% reconstruct the missing page 19
%%% from the original manuscript:
\frac{N}{2\pi}*2Re \left\{\sum_{L=0}^{\frac{N}{2}-1}c_L^i
	\exp \left(\frac{iL\cdot I2\pi}{N}\right)\right\}.
\end{align}
Now we use the identifications 
(\ref{idsvc17}) and comparing the 
coefficients we obtain
first for $n=L\ne 0$
\begin{equation}
\sqrt{2\alpha^{\prime}} (\alpha_n^{\mu}+
\alpha_{-n}^{\mu}) =\frac{N}{2\pi}(c_n^{\mu}+ c_n^{\mu \dagger}). 
\end{equation}
By nearer contemplation indeed we can obtain 
\begin{equation}
\alpha_n^{\mu}=
\frac{N}{2\pi\sqrt{2\alpha^{\prime}}}
c_n^{\mu}  \label{csvc19}
\end{equation}
 
%Similarly we deduce by identifying the 
%$\dot{X}_R(\tau_R(I)) = 
%\frac{N}{2\pi}J_R^{\mu}(I)$ for odd $I$ 
%we 
%obtain  a relation 
%of the $\alpha_n^{\mu}$ to the 
%$d_n^{\mu}$'s:\begin{equation}
%\sqrt{2\alpha^{\prime}} ()=...
%\end{equation} 
  
\setcounter{equation}{31}
This is
true for smoothly lying strings because 
we ``swindled" a bit by only using the 
even $I$.

We could instead have used the odd $I$'s 
still assuming smoothness and then we 
can express the relation between the 
$d_L^i=d_n^i$'s of the cyclically ordered chain and the $\alpha_n^i$'s. Let us in fact first insert (\ref{pidsvc13}) into the formula 
\begin{align}
	J^i(I)=-\pi \alpha^{\prime}\left\{\Pi^i(I+1)-\Pi^i(I-1)\right\}
\end{align}
from 
%\verb|
\cite{early2}
%|
for the odd $I$ ``objects" so as to obtain 
\begin{align}
	J^i(I)=-2\pi \alpha^{\prime}Re 
\left[
	\sum_{L=0}^{\frac{N}{2}-1}d_L^i\left\{
	\exp \left(\frac{iL(I+1)2\pi}{N}\right)
	-\exp \left(\frac{iL(I-1)2\pi}{N}\right)\right\}\right]
\end{align}
which in turn like just above is 
\begin{align}
	J^i(I) & \simeq 
\frac{2\pi}{N}\dot{X}_R^i
\left(\frac{I2\pi}{N}\right) 
	\nonumber \\
	& \simeq -\alpha^{\prime}
Re \left(2i\sum_{n=1}^{\infty}
	nd_n^ie^{\frac{inI\cdot 2\pi}{N}}
\right).
\end{align}
Using $\alpha_{-n}^i=\alpha_n^{i\dagger}$ we 
can write the left hand side as 
\begin{align}
	\alpha^{\prime}p^i+
\sqrt{2\alpha^{\prime}}\> 2Re 
	\left(\sum_{n=1}^{\infty}\alpha_n^ie^{-in(\tau -\sigma)}\right)
\end{align}
and we deduce 
\begin{align}
	-\alpha^{\prime}2ind_n^i=\sqrt{2\alpha^{\prime}}\> 2\alpha_n^i
\end{align}
using the identification (\ref{idsvc17}): 
\begin{align}
	e^{-in(\tau -\sigma)}=e^{i\frac{2\pi In}{N}}.
\end{align}
So we obtain by comparison: 
\begin{align}
	\alpha^{\prime}p^i & +\sqrt{2\alpha^{\prime}}
	\sum_{\genfrac{}{}{0pt}{}{n\neq 0}{n=-\infty}}^{+\infty}
	\alpha_n^ie^{-in(\tau -\sigma)} \nonumber \\
	& =-2\pi \alpha^{\prime}\cdot \frac{N}{2\pi}Re \left[
	\sum_{n=1}^{\frac{N}{2}-1}d_n^i
	\left\{\exp \left(\frac{in(I+1)\cdot 2\pi}{N}\right)
	-\exp \left(\frac{in(I-1)\cdot 2\pi}{N}\right)\right\}\right] 
	\nonumber \\
	& \simeq -\alpha^{\prime}NRe \left[2i\sum_{n=1}^{\frac{N}{2}-1}
	d_n^i\sin \left(\frac{n}{N}2\pi \right)e^{\frac{inI2\pi}{N}}\right].
\end{align}
Simplifying we get 
\begin{align}
	d_n^i=+i\sqrt{\frac{2}{\alpha^{\prime}}}\frac{\alpha_n^i}{n}.
	\label{dsvc22}
\end{align}
In 
%\verb|
\cite{ourappear}
%|\textcolor{red}{$\Leftarrow$(undefined!!)%} 
as formula (90) there had a smoothness 
requirement 
\begin{align}
	c_L^i\simeq 
-i4\pi\cdot \frac{\alpha^{\prime}}{N}d_L^iL
\end{align}
which with $n=L$ reads 
\begin{align}
	c_n^i\simeq 
-i\frac{4\pi \alpha^{\prime}}{N}d_n^in.
\end{align}
It is a self-consistency check that our 
two equations (\ref{csvc19}) and 
%\verb|
(\ref{dsvc22})
%|
\begin{align}
	& c_n^i=\frac{2\pi 
\sqrt{2\alpha^{\prime}}}{N}\alpha_n^i \\
	& d_n^i=
i\sqrt{\frac{2}{\alpha^{\prime}}}
\frac{\alpha_n^i}{n}
\end{align}
leading to 
\begin{align}
	c_n^i=
\frac{i2\pi\alpha^{\prime}n}{N}d_n^i
\end{align}
(which apart from a factor $2$ is the continuity condition). We should -- if we consider the Hilbert spaces of the chain of objects and the string as different -- really not have written our identification equations as we just did but rather have included an $F$ so as to rather be e.g. 
\begin{align}
	c_n^i=\frac{2\pi \sqrt{2\alpha^{\prime}}}{N}F^{-1}\alpha_n^iF
\end{align}
and 
\begin{align}
	d_n^i=i\sqrt{\frac{2}{\alpha^{\prime}}}F^{-1}\frac{\alpha_n^i}{n}F.
\end{align}

%%%%%%%%%%%%%%%%%%%%%%%%%%%%%%%%%%%%%%%%%%%%%%%%%%%%%%%%%%%%%%%%%%%%%%%%%%%%%%%
\section{String unification}
\label{unification}
\setcounter{equation}{46}

Now the main point of our string field theory in terms of the (actually even) ``objects" is that e.g. a splitting of an open string into two open strings in our model just reflects that one cyclically ordered chain ``after the decay" get interpreted as two such chains. That is to say ``ontologically" nothing happens, it is only that we ``after the decay" \underline{think} of two chains. Now we shall make the assumption that even though there are big quantum fluctuations there is still some element of continuity in the string and thus also in our cyclically ordered chains, so that the variation with $I$, odd or even, of $J^{\mu}(I)$ is at least somewhat continuous (in a crude way). Such a continuity if it is implemented statistically in the type of cyclically ordered chains which we consider, will mean that the probability for two different pictures of a system of objects in terms of chains will be more likely to be able to describe  some object-situation the more the 
hanging together in the two chain systems 
are the same. Let us attempt to say that 
in more detail: A certain set of even 
``objects" can be arranged into chains in 
many ways of course. See for instance on 
the figure here of say 18 even ``objects" 
can be put on cyclically ordered chains like this.
\begin{figure}[!htb]
 \includegraphics[width=0.2\textwidth]{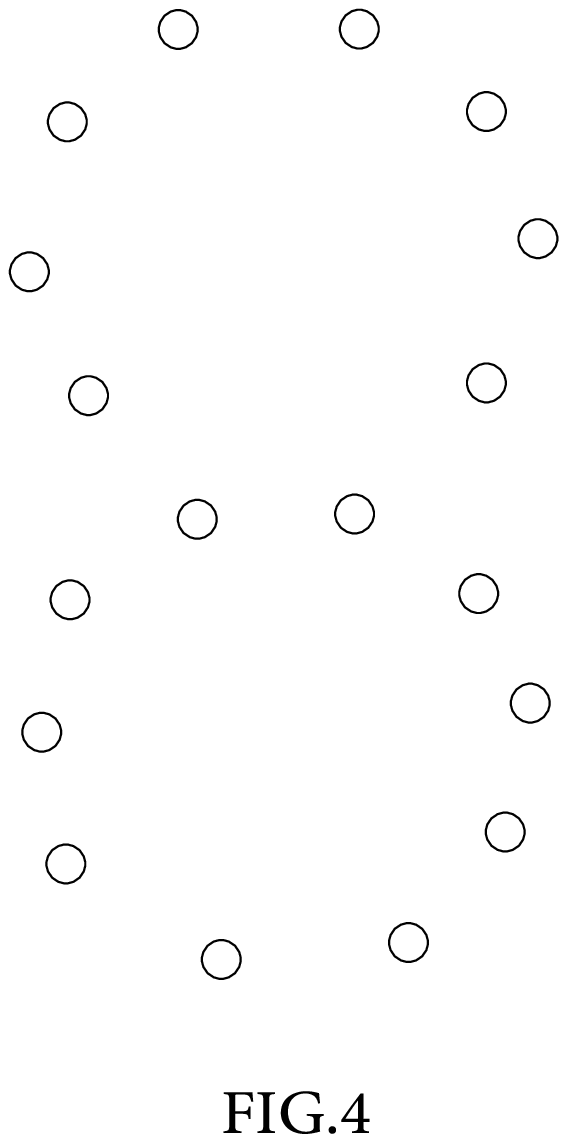}%
 \label{Figure4}
\end{figure}
\begin{figure}[!htb]
 \includegraphics[width=0.2\textwidth]{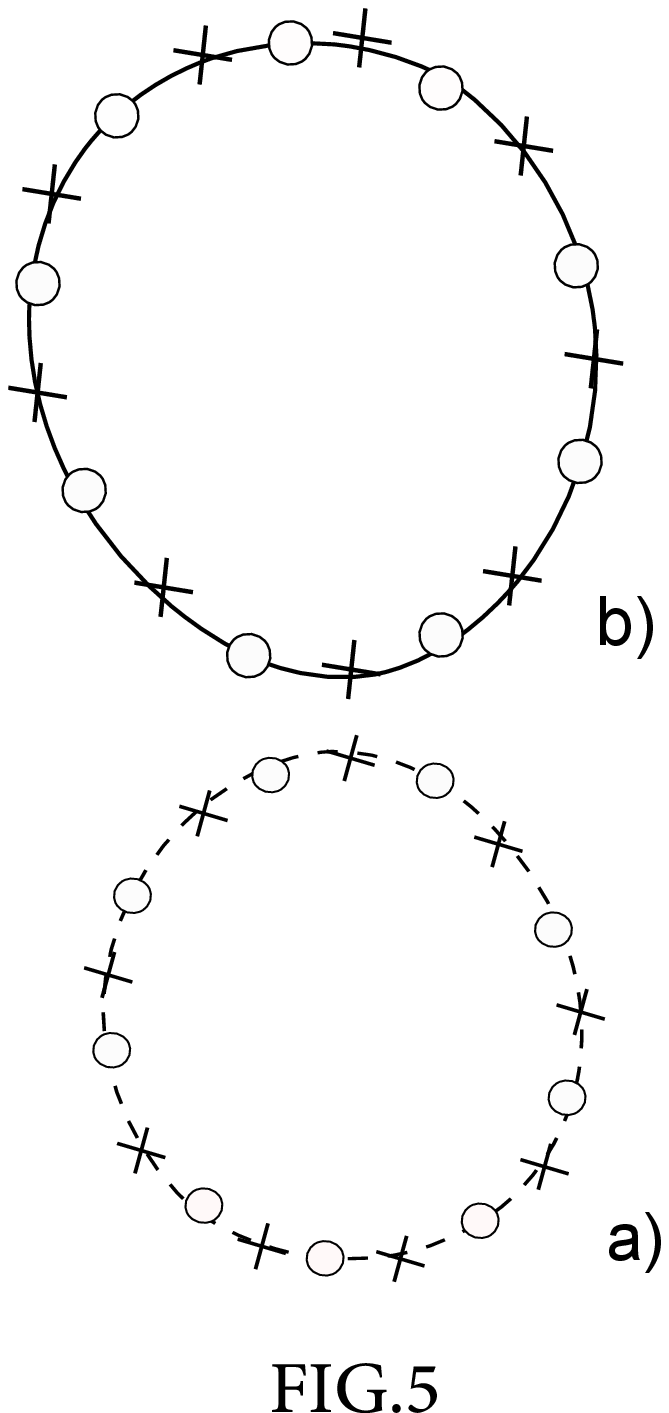}%
 \label{Figure5}
\end{figure}
\\
It is put on two chains. We can also put the same even objects, just on a single cyclically ordered chain like this. 
\begin{figure}[!htb]
 \includegraphics[width=0.23\textwidth]{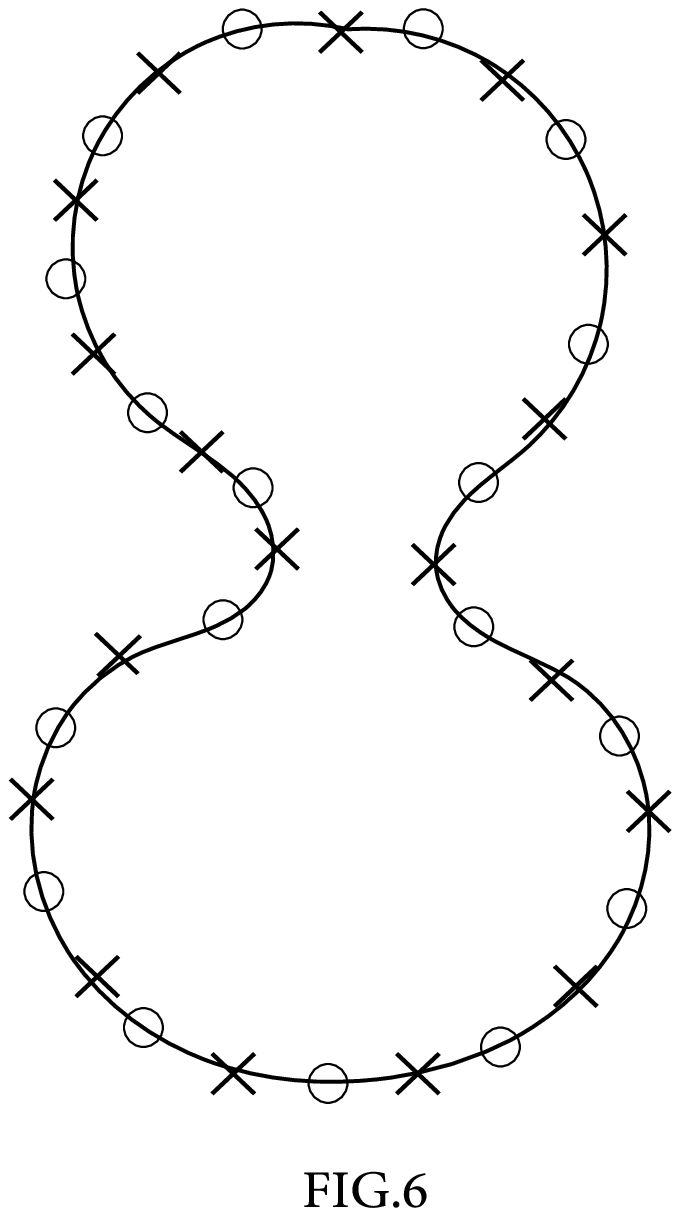}%
 \label{Figure6}
\end{figure}
\\
Now we could in principle take the ``objects" in any order along the chains, but we shall by the ``assumption of continuity" assume that the state of the even objects ``found in nature" is so that it may only be a few systems of chains of specific ordering that matches to have successive objects close in $J^i$ and $\Pi^i$ spaces to their neighbors in the cyclically ordered chain in which they sit. If we denote the ``objects" number $I$ in the chain number $\iota$ as $J^i(I,\iota)$ the $\Pi^i$ as $\Pi^i(I,\iota)$ we mean by continuity 
\begin{align}
	& J^i(I,\iota)\simeq J^i(I\pm 2,\iota), \\
	& \Pi^i(I,\iota)\simeq 
\Pi^i(I\pm 2,\iota).
\end{align}
Really we want even to assume such 
continuity even for odd ``objects" on the chains 
so that we have 
\begin{align}
	J^{\mu}(I,j)\simeq J^{\mu}(I\pm 1,j).
\end{align}
With such continuity the probability that two different sets/systems of cyclically ordered chains of objects can match -- with continuity -- to the some sets of ``objects" is of course biggest if most neighboring objects with respect to one set of chains are also neighbors with respect to the other one. This means that if e.g. a single cyclically ordered chain of objects under a ``decay" shall go to be in \underline{thought} replaced by say two cyclically ordered chains the continuity assumption makes it have highest probability to match the continuity if as many object-neighbors remain neighbors after the rethinking of the way the objects are put into chains. To make by rethinking one circular chain into two one needs topologically to violate the neighborship at two places at last on the single-starting chain. But since probability of matching with continuity is higher the lower the number of neighborship violations we suggest that in first approximation we shall think of rethinkings 
into a new chain-system which violate minimally the neighborship of the even ``objects".

Especially we shall think of the dominant rethought or ``decayed" single cyclically ordered chain into two chains as a change in the chaining consisting in cutting the initial chain at two places and choosing the two pieces to the two new chains. This is illustrated by the transition: 

\begin{figure}[!htb]
 \includegraphics[width=0.2\textwidth]{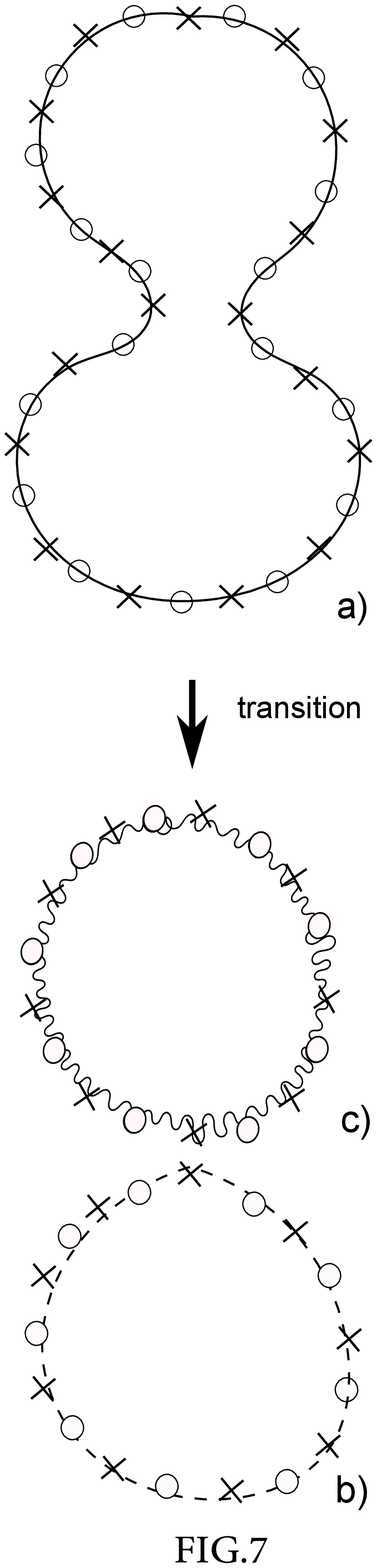}%
 \label{Figure7}
\end{figure}

To be understood that full line ---$\circ$---$\circ$---$\circ$---$\circ$--- chain gets split into the -- $\circ$-- $\circ$-- $\circ$-- $\circ$-- and the $\sim \circ \sim\circ\sim\circ\sim\circ\sim\circ\sim$ chains. This is done by only breaking the full curve ---$\circ$---$\circ$---$\circ$---$\circ$---$\circ$--- at two places.

The basic idea to obtain scattering amplitudes or vertices in our SFT consists in: 

\begin{enumerate}
\item[1)] Put up the state of the in general several strings (but in the case a string decay only one string) in the initial state and compute their state to be described in terms of ``objects" (really even objects at the end).

The reformulation from one string to one cyclically ordered chain is done by the formulas above. The composition of several strings $j=1,2,\cdots ,m$ to a SFT state with $m$ strings is in our picture simply done by considering the union of the sets of even objects associated with the strings in the initial state.

It becomes a state with 

\begin{align}
	\frac{1}{2}\left(N_1+N_2+\cdots +N_{\iota}+\cdots +N_m\right)
\end{align}

even objects if there are $\frac{1}{2}N_{\iota}$ even objects in to the $\iota$-th string corresponding cyclically ordered chain.
\item[2)] One then considers a similar final state with some other number $m_f$ of strings in it. From that one constructs quite analogously to the just described initial state a state consisting of $m_f$ chains of ``objects".
\item[3)] One takes the quantum mechanical (Hilbert inner product) overlap of the initial state even ``object" wave function $\psi_i$ and the final state even object wave function $\psi_f$ 

\begin{align}
	\int \psi_f^{\ast}(J)\psi_i(J)\prod_{I,\iota}\prod_{i=1}^{24}dJ^i(I,\iota).
	\label{osvc42}
\end{align}

For this overlap to make sense it is crucial that there is a (natural) matching between the even objects in the initial state with 

\begin{align}
	J^i(I_{\iota},{\iota})
\end{align}

where $J=1,2,\cdots ,m(=m_i)$ and $I_{\iota}=0,2,\cdots ,N_{\iota}-2$ and in the final state where we have $J^i(I_{\iota},\iota)$ but now with $\iota =1,2,\cdots ,m_f$ 
and $I_{\iota}=0,2,\cdots ,N_{\iota}^{(f)}-2$. Especially it is of course for common overlap (\ref{osvc42}) to be meaningful at least needed that total number of even objects in the initial state and in the final state is the same, i.e. 

\begin{align}
	N_1^{(i)}+N_2^{(i)}+\cdots +N_{\iota}^{(i)}+\cdots +N_{m_i}^{(i)}
	=N_1^{(f)}+N_2^{(f)}+\cdots +N_{\iota}^{(f)}+\cdots +N_{m_f}^{(f)}.
\end{align}
In the case we use the ``gauge condition" in IMF (=infinite momentum frame) that 
\begin{align}
	J^+(I_j,\iota)=\frac{a\alpha^{\prime}}{2} \quad 
	\text{(for both even and odd $I_{\iota}$)}
\end{align}
for all the objects the equality of the numbers of even objects follow from the conservation of $p^+$ (= the longitudinal momentum) and the fact that there is ``all the time (i.e. in all chainings) the same numbers of even and odd ``objects". So the conservation of the longitudinal momentum $p^+\propto \sum_{\text{all the objects}}J^+$ ensures that the number of objects in $\psi_i$ and $\psi_f$ match.
\end{enumerate}

%%%%%%%%%%%%%%%%%%%%%%%%%%%%%%%%%%%%%%%%%%%%%%%%%%%%%%%%%%%%%%%%%%%%%%%%%%%%%%%
\subsection{The trick of taking One String with Short Chain}
\label{trickshort}
In principle the reader should now that one should now calculate the vertex of the decay of one string into two by evaluating the overlap of two states for a certain number of even objects (proportional to $p^+$) calculated respectively from the single cyclically ordered chain of the to decay open string and from the unification/product of the two cyclically ordered chains corresponding to decay-product strings.

Now it is our hope to derive that vertex 
coming out of such an overlap calculation 
between a two string state and a 
one-string state should be calculated to 
be the three string vertex.

Now, however, the technically simplest 
vertex presented in string theory is 
rather the vertex for a string in an 
arbitrary state emitting a tachyon 
(ground state) string and becoming some 
to be prescribed state of the string. In 
such a vertex we can in principle express 
the vertex by a vertex operator such 
as (2.2.59) in \cite{GSW} 
\begin{align}
	V(k,\tau)=:e^{ik\cdot X(0,\tau)}:.
\end{align}
Notice that such a nice vertex operator 
is at least without the normal ordering $:\cdots :$ local on the string. It consists 
alone of string operators at the point 
$\sigma =0$.

In our picture wherein a priori a finite 
part of the cyclically ordered chain 
breaks off -- and therefore also a finite 
nonzero part of the string -- it would 
be very strange physically, if that could 
be described by an in $\sigma$ localized 
vertex operator. We would rather expect a 
non-local operator in $\sigma$ with an 
extent of the order of the string piece 
breaking off. That means that, if we 
insist on  going  for a vertex with an in $\sigma$ local vertex operator, then 
intuitively we would expect that the 
string piece breaking off or the piece of 
cyclically ordered chain breaking off 
should be so short that its length say in 
$\sigma$ goes to zero.

Let us take this way of thinking and 
this type of vertex operator, to be the 
one we  hope 
to achieve, as the suggestion to seek to 
make the breaking off piece infinitely 
small.

Such an infinitely small cyclically 
ordered chain of ``objects'' means also 
that its plus-component $p^+$ is very 
small 
and thus in our ``gauge''( ) having fixed 
to a small quantity  $\propto a$ the 
number of ``objects'' in the ``small''
piece is {\em relatively} very small
(but it can still be very large in 
absolute number). 

If one or more of the say transverse 
components for separating off string piece
is finite, then these components will
be relatively very big for the ``objects''
in this piece compared to the ones in the 
rest of the cyclically ordered chains.
This in turn means that in the continuum 
formulation as a density in $\sigma$
there is a delta-function contribution
from the external string represented by 
the ``small piece''. 

The achievement of this special choice of 
arranging the ``small piece'' to have 
small plus component is that we can use 
{\em the same parametrization for the 
two other external strings in the vertex 
under construction}. This is very 
important for obtaining a formalism like 
the one used in the usual formulation of 
the vertex operator, in which the ``small
piece'' gets represented by an operator,
while the two other external strings are 
represented by respectively the ket and 
the bra.  

Looking at a not infinitesimal scale in 
the $\sigma$ the breaking off piece is 
infinitesimally short. Nevertheless a
finite amount of say $p^i$ momentum 
is added to the incoming string in order 
to bring it to the ``outgoing one'' after 
the ``absorption '' of the counted as 
short string. This addition only 
comes in in the very small $\sigma$
region where the ``short string'' gets 
included. To describe the insertion 
of the ``short string'' (or short 
cyclically ordered piece of chain 
of ``objects'') in terms of an operator
we must then have an operator that changes
the momentum by the amount $k^{\mu}$ -
the momentum of the string related to the 
short cyclically ordered chain. Such an 
operator is precisely 
\begin{equation}
\exp(ik\cdot X(0,\tau)).
\end{equation}  
Apart from an overall constant factor and apart
from derivatives of delta-functions
this is just the usual vertex. So in this
sense we can claim that  we derived the 
usual vertex in our case considered.

Our derivation was a bit quick at least
with respect  to that we used that we 
could choose the frame we wanted 
but assuming that the theory of ours is
Lorentz invariance. That we hope but 
it is actually highly nontrivial.

%%%%%%%%%%%%%%%%%%%%%%%%%%%%%%%%%%%%%%%%%%%%%%%%%%%%%%%%%%%%%%%%%%%%%%%%%%%%%%%
%\renewcommand{\thesection}{}
\section{Conclusion and Resume}
\label{conclusion}

\begin{acknowledgments}
The authors acknowledge K. Murakami for informing them some useful information of the references on string field theory.
One of us (H. B. N.) thanks the Niels Bohr Institute for allowance to work as emeritus and for Masao to visit to give a talk in connection with the ``Holger Fest''. Masao Ninomiya acknowledges the Niels Bohr Institute and the Niels Bohr International Academy for inviting to the ``Holger Fest''. He also acknowledges that the present research is supported in part by the J.S.P.S. Grant-in-Aid for Scientific Research Nos.21540290, 23540332 and 24540293. Also H. B. N. thanks to the Bled conference participants organizers and Matijaz Breskov for financial support to come there where many of the ideas of this work got tested again.
\end{acknowledgments}


\begin{thebibliography}{99}
%%%%%%%%%%%%%%%%%%%%%%%%%%%%%%%%%%%%%%%%%%%%%%%%%%%%%%%%%%%%%%%%%%%%%%%%%%%%%%%
%\bibitem{rf:gsw}
%M. B. Green, J. H. Schwarz \& E. Witten, `%`Superstring theory Volume 1: Introduction%", Cambridge University Press.
%%%%%%%%%%%%%%%%%%%%%%%%%%%%%%%%%%%%%%%%%%%%%%%%%%%%%%%%%%%%%%%%%%%%%%%%%%%%%%%
%\end{thebibliography}
%%%%%%%%%%%%%%%%%%%%%%%%%%%%%%%%%%%%%%%%%%%%%%%%%%%%%%%%%%%%%%%%%%%%%%%%%%%%%%%




\bibitem{early2}
%[1] 
H. B. Nielsen and M. Ninomiya ``An Idea of New String Field Theory - Liberating Right and Left movers - '',
 in Proceedings of the 14th Workshop, ``What Comes Beyond the Standard Models'' Bled, July 11-21,
2011, eds. N. M. Borstnik, H. B. Nielsen and D. Luckman; ArXiv: 1112.542 [hep-th].

\bibitem{early1}
%[2]
H. B. Nielsen and M. Ninomiya ``A New Type of String Field Theory'', in Proceedings of the 10th
Tohwa International Symposium in String Theory, July 3-7, 2011, Fukuoka Japan, AIP conference
Proc. vol.607, p.185-201; ArXiv: hep-th/0111240.

\bibitem{ourappear}
%[22] 
H. B. Nielsen and M. Ninomiya,``A Novel String Field Theory Solving String Theory by Liberating Left and Right Movers'',
ArXiv: 1211.1454 [hep-th].

\bibitem{lcgbosonic}
%[3] 
As for bosonic string field theory in the light-cone gauge:
M. Kaku and K. Kikkawa, Phys. Rev. D10(1974)1110; M. Kaku and K. Kikkawa, Phys. Rev.
D10(1974)1823; S. Mandelstam, Nucl. Phys. B64(1973)205; E. Cremmer and J.-L. Gervais,
Nucl. Phys. B90(1975)410.
\bibitem{WittenSFT}
%[4] 
Witten type mid-point interaction of covariant bosonic string field theory: for open string
E. Witten, Nucl. Phys. B268(1986)253.
\bibitem{closedWitten}
%[5] 
Closed string of the Witten theory:
M. Saadi and B. Zwiebach, Ann. Phys. 192(1989)213; T. Kugo, H. Kunitomo and K. Suehiro,
Phys. Lett. B226(1989)48; T. Kugo and K. Suehiro, Nucl. Phys. B337(1990)434; B. Zwiebach,
Nucl. Phys. B390(1993)33.
\bibitem{JSWittenopen}
%[6] 
Joining-Splitting light-cone type of Witten theory: for open string
H. Hata, K. Itoh, T. Kugo, H. Kunitomo and K. Ogawa, Phys. Rev. D34(1986)2360.
\bibitem{JSWittenclosed}
%[7] 
Closed string of Witten theory of Joining-Splitting light-cone type:
H. Hata, K. Itoh, H. Kunitomo and K. Ogawa, Phys. Rev. D35(1987)1318; H. Hata, K. Itoh,
T. Kugo, H. Kunitomo and K. Ogawa.
\bibitem{susyWittenopen}
%[8] 
Supersymmetrized covariant open Witten theory
E. Witten, Nucl. Phys. B276(1986)291;
B212(1988)299;
C. Wendt,
I. Arefeva and P. Medvedev, Phys. Lett.
Nucl. Phys.B314(1989)209;
N. Berkovits,
Nucl. Phys.
B450(1995)90 [Errata, B459(1996)439] hepth/9503099.
\bibitem{modifiedcubicWitten}[9] Modified cubic Witten Theory:
I. Arefeva, P. Medvedev and A. Zubarev, Nucl. Phys. B341(1990)464; C. R. Preitschopf,
34
C. B. Thorn and S. A. Yost, Nucl. Phys. B337(1990)363.
\bibitem{susyGS}
%[10] 
Supersymmetrized Green-Schwarz theory in the light-cone gauge:
M. B. Green and J. H. Schwarz, Nucl. Phys. B218(1983)43; M. B. Green, J. H. Schwarz
and L. Brink, Nucl. Phys. B219(1983)437; M. B. Green, and J. H. Schwarz, Nucl. Phys.
B243(1984)437; J. Greensite and F. R. Klinkhamer, Nucl. Phys. B281(1987)269; Nucl.
Phys. B291(1987)557; Nucl. Phys. B304(1988)108; M. B. Green and N. Seiberg, Nucl. Phys.
B299(1988)559; S.-J. Sin, Nucl. Phys. B313(1989)165.

\bibitem{reviewSFT}
%[11] 
For review articles of string field theory: W. Taylor and B. Zwiebach, D-branes, Tachyons, and String
Field Theory hep-th/0311017 http:jp.ArXiv.org/abs/hep-th/0311017; K. Ohmori, A Review on Tachyon
Condensation in Open String Theories hep-th/0102085 http://jp.ArXiv.org/abs/hep-th/0102085.

\bibitem{stringNielsen}
%[12] 
String references H. B. Nielsen A Physical interpretation of the integrand of the n-point Veneziano model
(1969)(Nordita preprint...); a later version An almost Physical interpretation of the integrand of the
n-point Veneziano model preprint at Niels Bohr Institute; a paper presented at the 15th International
Conference on High Energy Physics, Kiev 1970(see p445 in Venezianofs talk.)

\bibitem{stringNambu1}
%[13] 
Y. Nambu, Quark Model and the factorization of the Veneziano amplitude, in Proceedings of the
International Conference on Symmetries and Quark Models, June 18-20, 1969 ed. by Chand, R. (Gordon
and Breach), New York, 269-277; reprinted in Broken Symmetry Selected Papers of Y. Nambu eds. T.
Eguchi and K. Nishijima, (World Scientific, Singapore, 1995)258- 277.

\bibitem{stringNambu2}
%[14] 
Y. Nambu, Duality and hadrondynamics, Lecture notes prepared for Copenhagen summer school, 1970;
It is reproduced in Broken Symmetry, Selected Papers of Y. Nambu eds. T. Eguchi, and K. Nishijima,
(World Scientific, Singapore, 1995)280.

\bibitem{stringSusskind}
%[15] 
L. Susskind, Structure of hadrons implied by duality, Phys. Rev. D1(1970)1182-1186(1970): L.
Susskind, Dual symmetric theory of hadrons 1. Nuovo Cimento A69,457-496(1970).

\bibitem{Birth}
%[16] 
For many original papers of early string theory see ``The Birth of String Theory'' eds. by Andrea Cappeli,
Elena Castellani, Flippo Colomo and Paolo Di Vecchia, Cambridge University Press 2012. Cambridge,
New York, Melbourne.

\bibitem{GSW}
%[17] 
M. B. Green, J. H. Schwarz and E. Witten Superstring theory, 1-2, Cambridge University
Press 1987.
\bibitem{Polchinski}
%[18] 
For comprehensive reviews of string theory see e.g. J. Polchinski String Theory vol.I-II, Cambridge
University Press 1998.

\bibitem{doubling}
%[19] 
Appearance of doubling, ``Confinement of quarks'', Phys. Rev. D10 2445(1974), see also ``Absence of 
neutrinos on a lattice I - Proof by homotopy theory - '' H. B. Nielsen and M. Ninomiya.

\bibitem{GGRT}
%[20] 
P. Goddard, J. Goldstone, C. Rebbi, C. B. Thorn, Nucl. Phys. B56 109(1973).

\bibitem{Veneziano}
%[21] 
``Construction of a crossing-symmetric, Regge behaved amplitude for linearly rising trajectories'', G.
Veneziano, Nuovo Cimento A57 190(1968).

%\bibitem{ourappear}
%[22] 
%H. B. Nielsen and M. 
%Ninomiya,``A novel ...'' to appear.
\bibitem{Maldazena}
%[23] 
``The Large N limit of superconformal field theories and supergravity'', J. M. Maldacena, Advances in
Theoretical and Mathematical Physics 2, 231(1998).
\end{thebibliography}
\end{document}